\documentclass[12pt]{article}
\usepackage[left=1.25in,top=1.5in,right=1.25in,bottom=1.25in]{geometry}
\usepackage{amsmath}
\usepackage{amssymb}
\usepackage{amsbsy}
\usepackage{amsthm}
\usepackage{epsfig}
\usepackage[round]{natbib}
\usepackage{clrscode}
\usepackage{setspace}
\usepackage{multirow}

\newcommand{\N}{\mbox{N}}

\newcommand{\C}{\mbox{C}}

\newcommand{\IG}{\mbox{IG}}
\newcommand{\Ga}{\mbox{Ga}}

\newcommand{\IB}{\mbox{IB}}

\newcommand{\bbeta}{\boldsymbol{\beta}}
\newcommand{\btheta}{\boldsymbol{\theta}}

\title{Parameter expansion in local-shrinkage models}

\author{
\textsc{James G.~Scott} \\
\small{\textit{McCombs School of Business}} \\ \small{\textit{University of Texas at Austin}} \\
}

\date{September 2010}
\begin{document}

\maketitle
\begin{abstract}

This paper considers the problem of using MCMC to fit sparse Bayesian models based on normal scale-mixture priors.  Examples of this framework include the Bayesian LASSO and the horseshoe prior.  We study the usefulness of parameter expansion (PX) for improving convergence in such models, which is notoriously slow when the global variance component is near zero.  Our conclusion is that parameter expansion does improve matters in LASSO-type models, but only modestly.  In most cases this improvement, while noticeable, is less than what might be expected, especially compared to the improvements that PX makes possible for models very similar to those considered here.  We give some examples, and we attempt to provide some intuition as to why this is so.  We also describe how slice sampling may be used to update the global variance component. In practice, this approach seems to perform almost as well as parameter expansion.  As a practical matter, however, it is perhaps best viewed not as a replacement for PX, but as a tool for expanding the class of models to which PX is applicable.

\vspace{0.1in}
\end{abstract}


\section{Parameter expansion for variance components}

\begin{figure}
\includegraphics[width=6.0in]{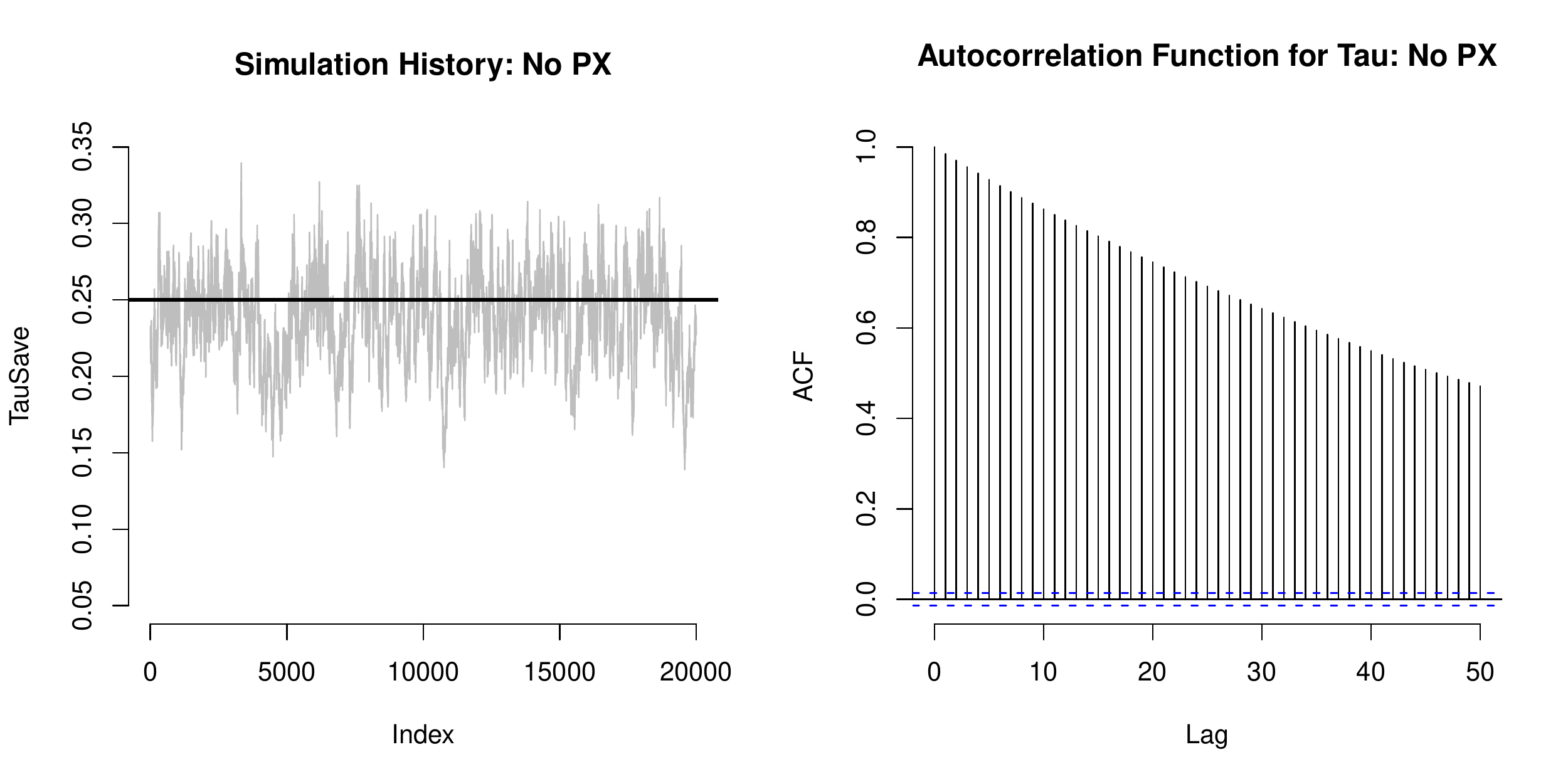}\\
\includegraphics[width=6.0in]{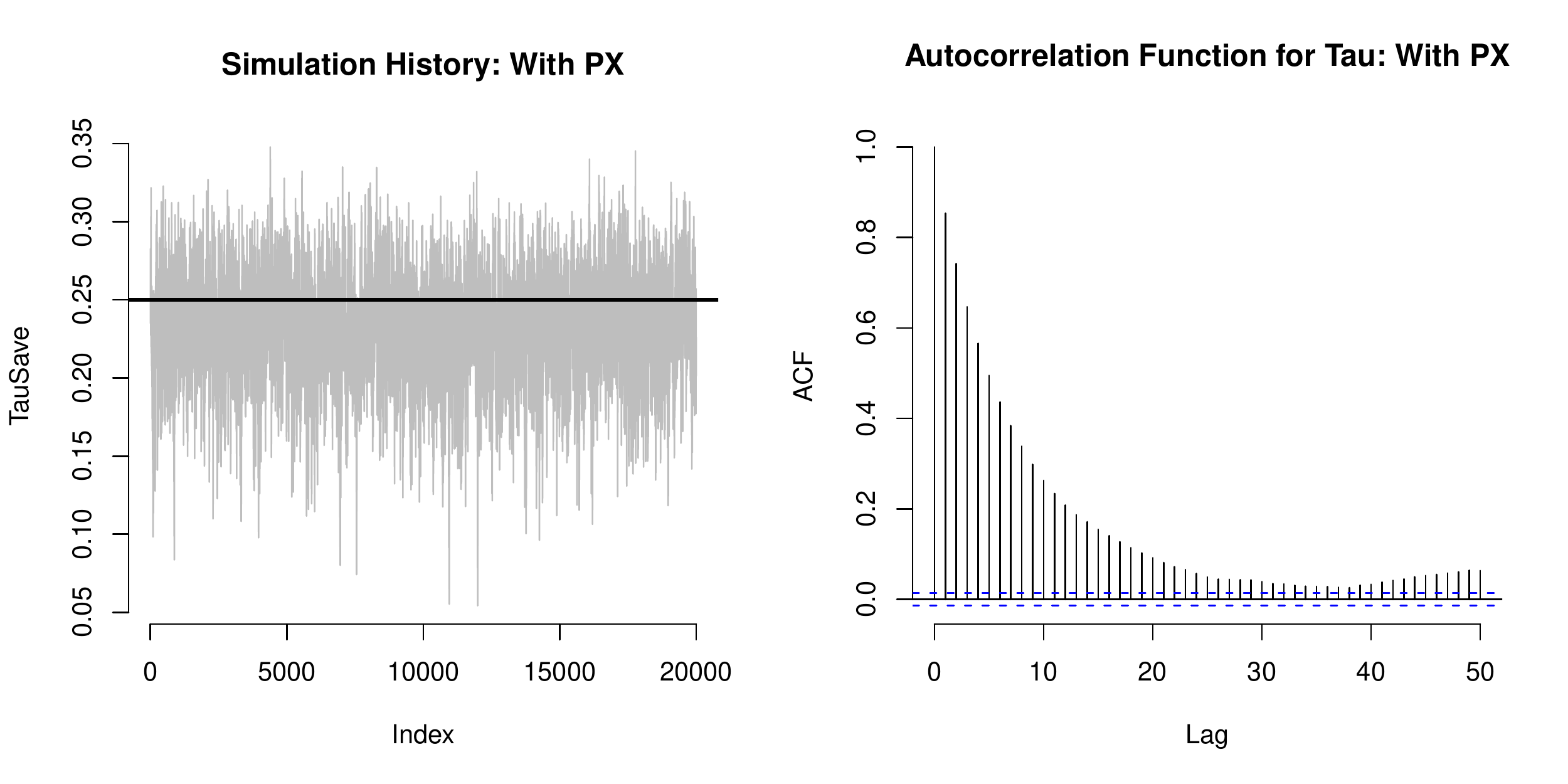}
\caption{\label{fig:globalPX} Above: Simulation history and autocorrelation plot for $\tau$ in the non-parameter-expanded Gibbs sampler.  Below: the same plots for the parameter-expanded sampler.}
\end{figure}

\begin{figure}
\begin{center}
\includegraphics[width=5.0in]{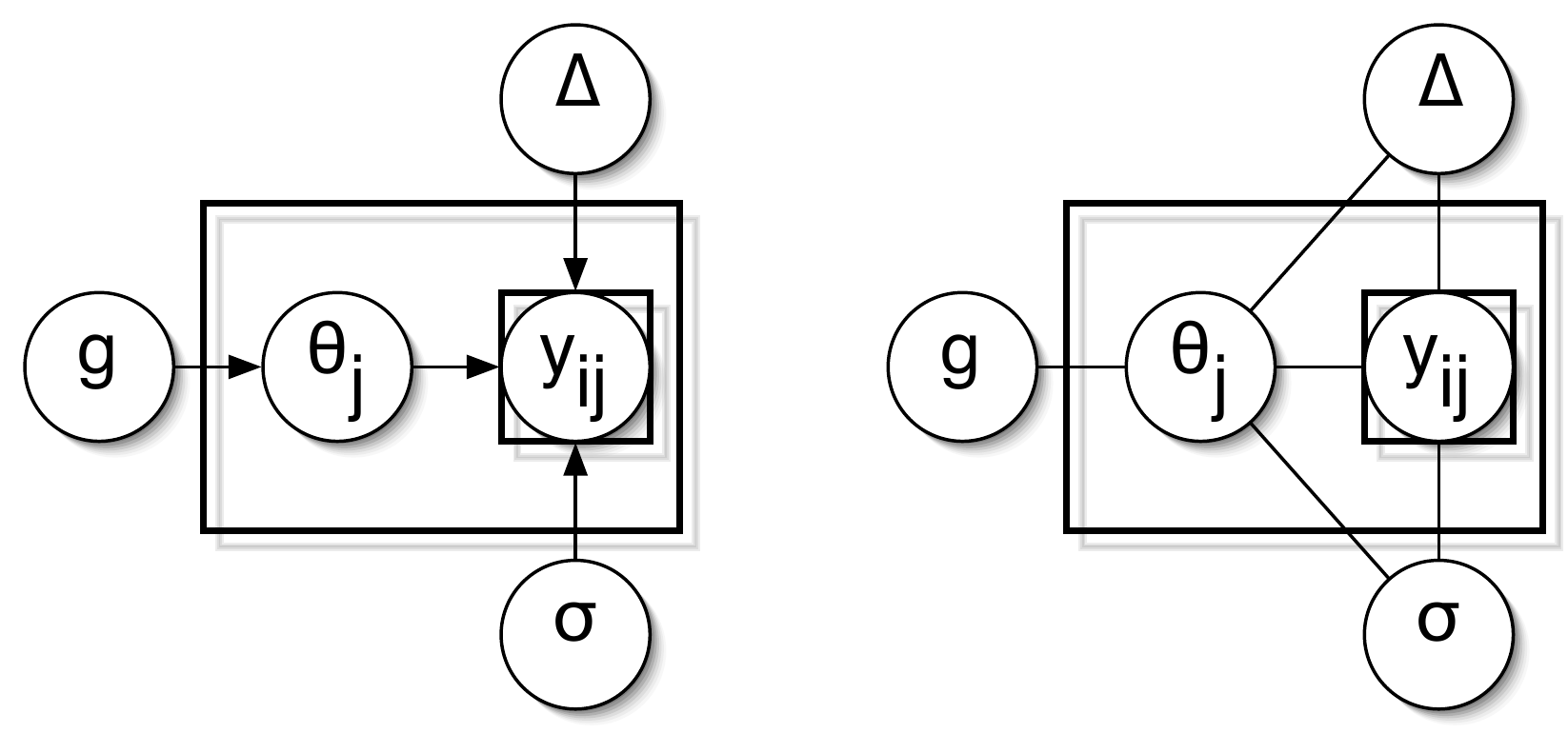}\\
\caption{\label{fig:globalPXgraph} The directed graph (left) and moralized undirected graph (right) corresponding to the PX model, (\ref{PXglobal1})--(\ref{PXglobal4}).  Circles indicate nodes; rectangles, replicates.}
\end{center}
\end{figure}

Many common Bayesian models have equivalent parameter-expanded (PX) versions, in which redundant, non-identified parameters are introduced for the sake of improving MCMC convergence.  In particular, we are interested in the following simple case and its generalizations:
\begin{eqnarray}
(y_{ij} \mid \beta_j, \sigma^2) &\sim& \N(\beta_j, \sigma^2) \label{noPXglobal1} \\
(\beta_j \mid \sigma^2, \tau^2) &\sim& \N(0, \sigma^2 \tau^2) \label{noPXglobal2} \\
\tau &\sim& \C^{+}(0,1) \label{noPXglobal3}
\end{eqnarray}
for $i = 1, \ldots, n$ and $j = 1, \ldots, p$.

The parameter-expanded (PX) model corresponding to (\ref{noPXglobal1})--(\ref{noPXglobal3}) is
\begin{eqnarray}
(y_{ij} \mid \beta_j, \sigma^2) &\sim& \N(\Delta \sigma \theta_j, \sigma^2) \label{PXglobal1}  \\
(\theta_j \mid g^2) &\sim& \N(0, g^2) \label{PXglobal2} \\
\Delta &\sim& \N(0,1) \label{PXglobal3} \\
g^2 &\sim& \IG(1/2,1/2) \label{PXglobal4} \, ,
\end{eqnarray}
In both cases, assume that $p(\sigma) \propto 1/\sigma$.  More generally, $\tau$ may have a positive noncentral-$t$ prior, and a similar equivalence will hold.  But the half-Cauchy prior is an excellent default choice for many problems \citep{gelman:2006,Polson:Scott:2010c}, and is a useful special case for the sake of illustration.

Despite the fact that $\Delta$ and $g^2$ are not individually identifiable, Model (\ref{PXglobal1})--(\ref{PXglobal4}) and Model (\ref{noPXglobal1})--(\ref{noPXglobal3}) are identical in all the important ways: the marginal likelihood in $y$, the model for $\beta_j$, and the implied prior for $\tau$.  It is easy, moreover, to translate between the two parameterizations, since $\beta_j \equiv \Delta \sigma \theta_j$ and $\tau \equiv |\Delta| g$.  Yet (\ref{PXglobal1})--(\ref{PXglobal4}) will result in an MCMC that converges faster---often drastically so, and especially when $\tau$ is close to zero.  This phenomenon has been widely studied, and has been exploited with great success to speed computation for many common Bayesian models \citep[see, e.g.,][for many useful references]{vandyk:meng:2001}.

For example, Figure \ref{fig:globalPX} compares the standard and PX Gibbs samplers for a particular simulated data set where $p=2000$, $n=3$, $\tau=0.25$, and $\sigma=1.25$.  The standard sampler exhibits severe autocorrelation, while the PX sampler appears healthy.  (R code for implementing all simulations can be found in Appendix \ref{app:globalcode}.)

The advantage of Model (\ref{PXglobal1})--(\ref{PXglobal4}) is apparent from Figure \ref{fig:globalPXgraph}, and arises from the fact that $\Delta$ and $g$ are conditionally independent in the posterior distribution, given $\btheta = (\theta_1, \ldots, \theta_p)'$.  The crucial fact is that $g$ enters the model at the level of the parameters, while $\Delta$ enters the model at the level of the data.  Since $\tau$ is the product of these two factors, each of which may vary independently in the conditional posterior distribution, the result is reduced autocorrelation.

\section{LASSO-type Bayesian models}

\subsection{Sparsity via scale mixtures}

\begin{figure}
\begin{center}
\includegraphics[width=5.0in]{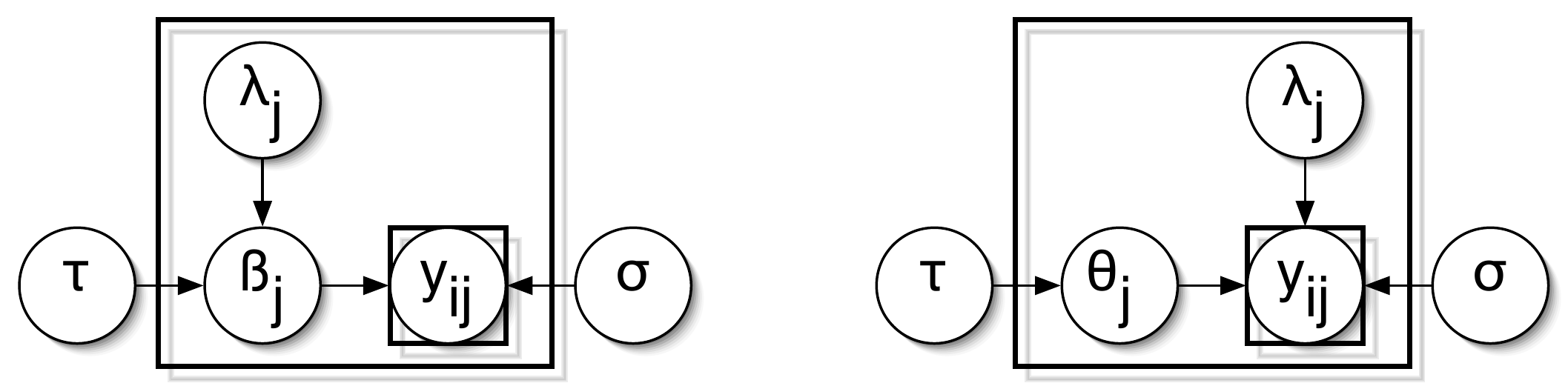}\\
\caption{\label{fig:localnoPXgraph} Two equivalent ways of expressing the non-PX local shrinkage model.}
\end{center}
\end{figure}

Many popular models for a sparse location vector $\bbeta = (\beta_1, \ldots, \beta_p)'$ assume an exchangeable prior for $\beta_j$, and can can be studied most readily in the generalization of (\ref{noPXglobal1})--(\ref{noPXglobal3}) to cases where $\bbeta$ is a sparse vector of normal means with a non-Gaussian prior.  We now consider the question of whether parameter expansion can offer improvements similar to those available in the pure Gaussian case.

Suppose that $(y_{ij} \mid \beta_j, \sigma^2) \sim N(\beta_j, \sigma^2)$ and the prior for $\beta_j$ is of the form $p(\beta_j/\{\sigma \tau\})$, where $\tau$ has a prior distribution.  Models of this form include the relevance vector machine of \citet{tipping:2001}; the double-exponential prior or ``Bayesian LASSO'' \citep{carlin:polson:1991,tibshirani1996,park:casella:2008,hans:2008,gramacy:pantaleo:2009}; the normal/Jeffreys prior \citep{figueiredo:2003, bae:mallick:2004}; the Strawderman--Berger prior \citep{strawderman:1971, BergerAnnals1980}; the normal/exponential/gamma prior \citep{griffin:brown:2005}; and the horseshoe prior \citep{Carvalho:Polson:Scott:2008a}, among many others.

As many authors have observed, autocorrelation for $\tau$ in these sparse models can cause great difficulty; see, for example, \citet{hans:2010}.  The logic of parameter expansion implies that we should instead allow two nonidentified parameters to perform the work of the single identified parameter $\tau$.  In light of the previous example, it is natural---but, as it turns out, na\"ive---to hope that this approach can solve the problem.

The difficulty is that all of these sparse models are best fit by introducing extra \textit{identified} parameters $\{\lambda^2_1, \ldots, \lambda^2_p\}$ to the model.  This is done in such as a way as to make the prior for $\beta_j$ conditionally Gaussian, given the local shrinkage factors $\lambda_j$.  This greatly simplifies the model.  But as we shall see, it also diminishes the potential advantage to be gained by introducing non-identified variance components.

To see this, write the local shrinkage generalization of (\ref{noPXglobal1})--(\ref{noPXglobal3}) as
\begin{eqnarray}
(y_{ij} \mid \beta_j, \sigma^2) &\sim& \N(\sigma \beta_j, \sigma^2) \label{noPXlocal1} \\
(\beta_j \mid \lambda_j^2, \tau^2) &\sim& \N(0, \tau^2 \lambda_j^2) \label{noPXlocal2} \\
\lambda_j &\sim& p(\lambda_j) \label{noPXlocal3} \\
\tau &\sim& \C^{+}(0,1) \label{noPXlocal4} \, ,
\end{eqnarray}
with Table \ref{tab:lambda.profiles} listing some choices for $p(\lambda_j)$ that give rise to common sparsity priors.

Alternatively, (\ref{noPXlocal1}) and (\ref{noPXlocal2}) may be re-written as
\begin{eqnarray}
(y_{ij} \mid \theta_j, \lambda_j, \sigma^2) &\sim& \N(\sigma \lambda_j \theta_j, \sigma^2) \label{noPXlocal1.2} \\
(\theta_j \mid \tau^2) &\sim& \N(0, \tau^2) \label{noPXlocal2.2}
\end{eqnarray}
with $\beta_j \equiv \lambda_j \theta_j$.  These two equivalent ways of writing the model are shown graphically in Figure \ref{fig:localnoPXgraph}.  Note the usual conjugate form preferred by \citet{jeffreys1961}, with the error variance $\sigma^2$ scaling the prior for the location vector.

\begin{table}
\begin{center}
\begin{small}
\caption{\label{tab:lambda.profiles} Priors for $\lambda_j$ associated with some common sparsity priors.  Densities are given up to constants and do not account for global scale terms.}
\vspace{1pc}
\begin{tabular}{l l }
Marginal prior for $\beta_j$ & Prior for $\lambda_j$ \\
\\
Double-exponential & $\lambda_j  \exp \left(-\lambda_j^2/2 \right)$  \\
Cauchy & $\lambda_j^{-2}  \exp\left\{ 1 / \left(2\lambda_j^2\right) \right\}$  \\
Strawderman--Berger & $\lambda_j \ (1 + \lambda_j^2)^{-3/2}$ \\
Normal--exponential--gamma &  $\lambda_j \ (1 + \lambda_j^2)^{-(c+1)}$  \\
Normal-Jeffreys & $\lambda_j^{-1}$  \\
Horseshoe & $(1+\lambda_j^2)^{-1} $ \\
\end{tabular}
\end{small}
\end{center}
\end{table}

\begin{figure}
\begin{center}
\includegraphics[width=5.0in]{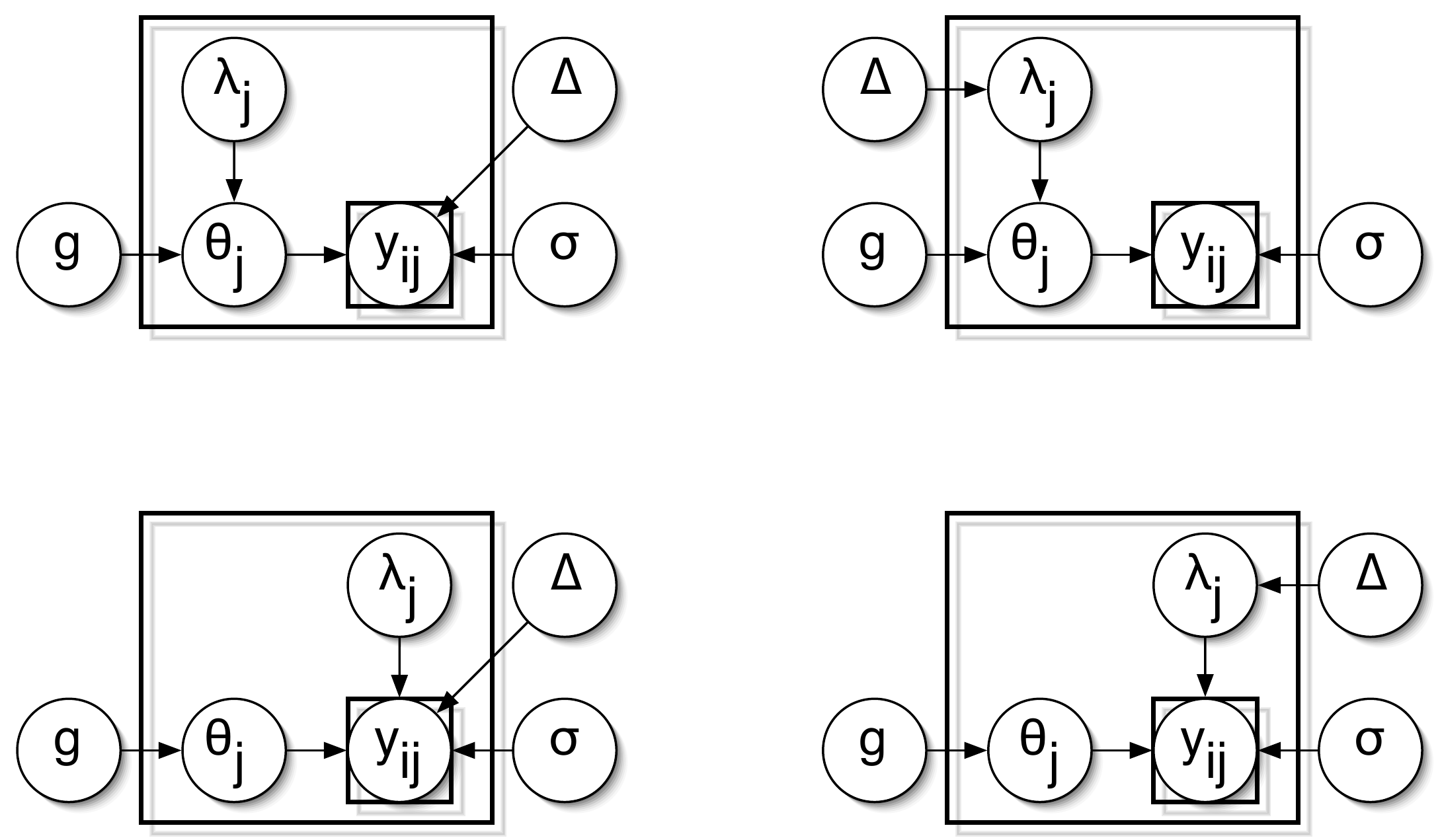}\\
\caption{\label{fig:localPXgraph} Four equivalent ways of expressing the parameter-expanded local shrinkage model.}
\end{center}
\end{figure}

The obvious PX approach is to let $\tau \equiv |\Delta| g$ as before, and to postprocess the MCMC draws for $g$ and $\Delta$ to estimate $\tau$.  For example, we might let
\begin{eqnarray}
(y_{ij} \mid \theta_j, \lambda_j, \Delta, \sigma^2) &\sim& \N(\sigma \Delta \lambda_j \theta_j, \sigma^2) \label{PXlocal1} \\
(\theta_j \mid g^2) &\sim& \N(0, g^2) \label{PXlocal2} \\
\lambda_j &\sim& p(\lambda_j) \label{PXlocal3} \\
\Delta &\sim& \N(0,1) \label{PXlocal4} \\
g^2 &\sim& \IG(1/2,1/2) \label{PXlocal5} \, ,
\end{eqnarray}
This version of the model, along with three equivalent versions, are shown graphically in Figure \ref{fig:localPXgraph}.  These versions differ in where $\Delta$ and $\lambda_j$ enter the hierarchy (that is, at the data level or parameter level), and correspond to different undirected graphs for the full joint distribution.

\subsection{The non-PX updates for $\tau$ and $\lambda_j$}

As an alternative to parameter expansion, we use the following approach based on slice sampling \cite[see, e.g.,][]{damien:etal:1999}.  Define $\eta_j = 1/\lambda_j^2$, and define $\mu_j = \beta_j/(\sigma \tau)$.  Then the conditional posterior distribution of $\eta_j$, given all other model parameters, looks like
$$
p(\eta_j \mid \tau,\sigma,\mu_j) \propto \exp \left\{ -\frac{\mu_j^2}{2} \eta_j \right\} \frac{1}{1+\eta_j} \, .
$$
Therefore, the following two steps are sufficient to sample $\lambda_j$:
\begin{enumerate}
\item Sample $(u_j \mid \eta_j)$ uniformly on the interval $(0, 1/(1 + \eta_j))$.
\item Sample $(\eta_j \mid \mu_j, u_j) \sim \mbox{Ex}(2/\mu_j^2)$ from an exponential density, truncated to have zero probability outside the interval $(0, (1-u_j)/u_j)$.
\end{enumerate}
Transforming back to the $\lambda$-scale will yield a draw from the desired conditional distribution.

The same trick works for $\tau$, letting $\eta = 1/\tau^2$ and replacing $\mu_j^2$ by $\sum \theta_j^2/2$.  Indeed, the approach will work for any prior for which the slice region in Step 1 is invertible, or can be transformed to make it invertible (as for the half-Cauchy).

Slice-sampling can also be used independently for $g$, $\Delta$, or both.  This tactic expands the class of variance-component priors to which parameter expansion is applicable.  For example, the noncentral positive $t$ distribution corresponds to $\tau = |\Delta| g$ where $\Delta \sim \N(m, 1)$ and $g^2 \sim \IG(a/2,b/2)$.  This leads to conditionally conjugate updates for both $\Delta$ and $g^2$.

Suppose, on the other hand, that one would prefer $\tau^2$ to have some prior other than a noncentral positive $t$.  Slice sampling makes this possible.  For example, let $\tau^2 \sim \IB(a,b)$, an inverted beta or ``beta-prime'' distribution.  This generalizes the half-Cauchy prior in a different direction, making $\tau^2$ equal in distribution to the ratio of two gamma random variables, or equivalently $\tau = |\Delta| g$ where $\Delta^2 \sim \Ga(a, 1)$ and $g^2 \sim \IG(b,1)$.  It is then possible to use the usual conjugate update for $g^2$ in the PX model, and to use slice sampling to update $\Delta^2$.  We do not explore this fact further, but note that it opens up the possibility of using PX to fit models involving even more general classes of variance-component priors.

All other draws are standard Gibbs updates and are omitted.

\section{Simulation results}

To compare the PX and non-PX samplers, we used the horseshoe prior of \citet{Carvalho:Polson:Scott:2008a}, where $\lambda_j \sim \C^{+}(0,1)$.  The resulting marginal prior distribution for $\beta_j$ has Cauchy-like tails and a pole at zero, and seems to perform very well as a default sparsity prior.

The line of reasoning behind the horseshoe prior is that $\tau$ should concentrate near zero \textit{a posteriori}.  This will provide a strong shrinkage effect for most observations (i.e.~the noise).  The signals, meanwhile, will correspond to very large values of $\lambda_j$, from far out in the tail of the half-Cauchy prior, allowing certain observations to escape the ``gravitational pull'' of $\tau$ toward zero.

Because it exhibits a striking antagonism between modeling goals and computational goals, the horseshoe prior makes for an interesting test case.  Convergence problems arise precisely when $\tau$ is small.  Yet the logic of the model says that $\tau$ must be small in order to squelch noise.

\begin{figure}
\includegraphics[width=6.0in]{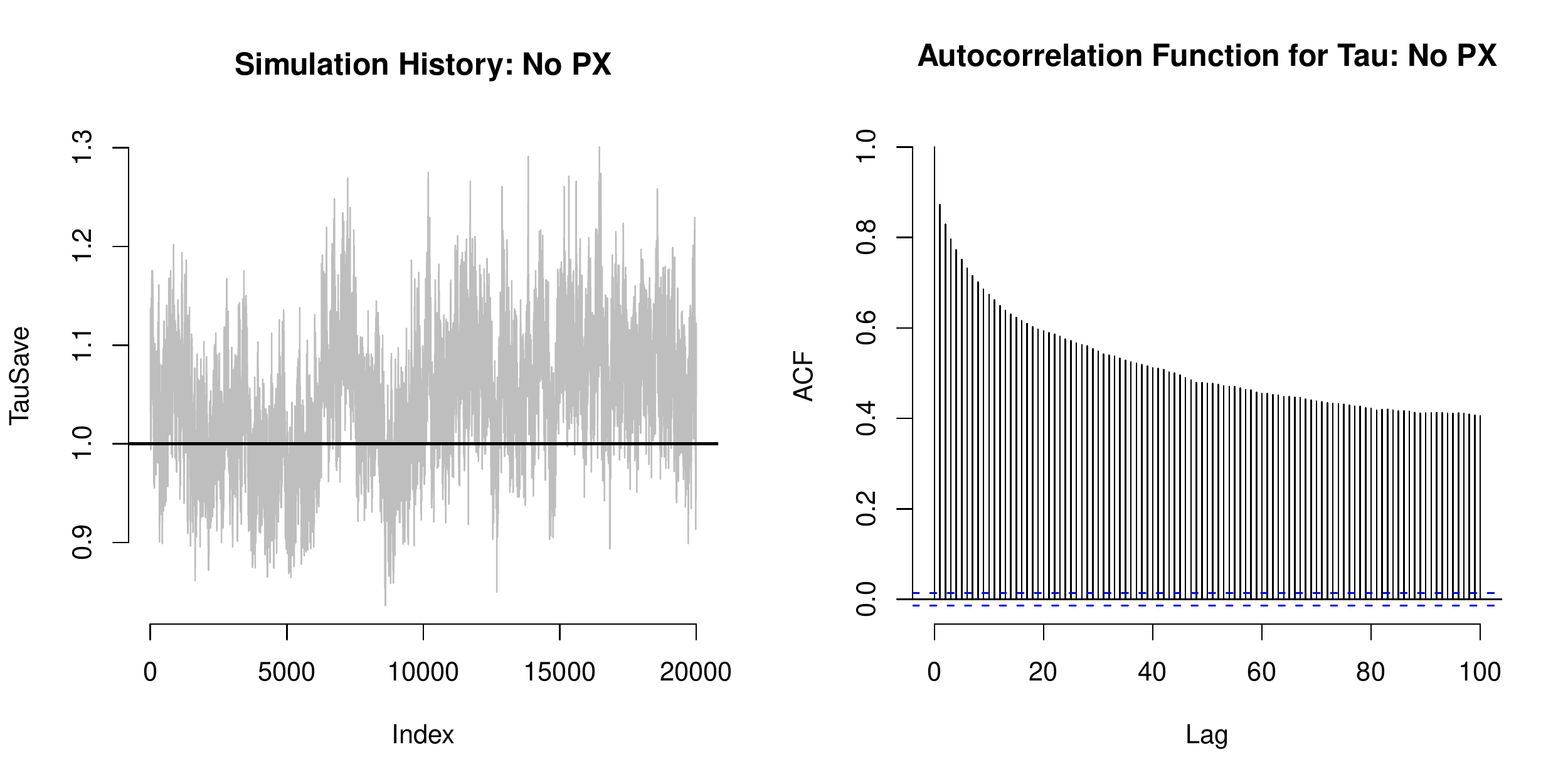}\\
\includegraphics[width=6.0in]{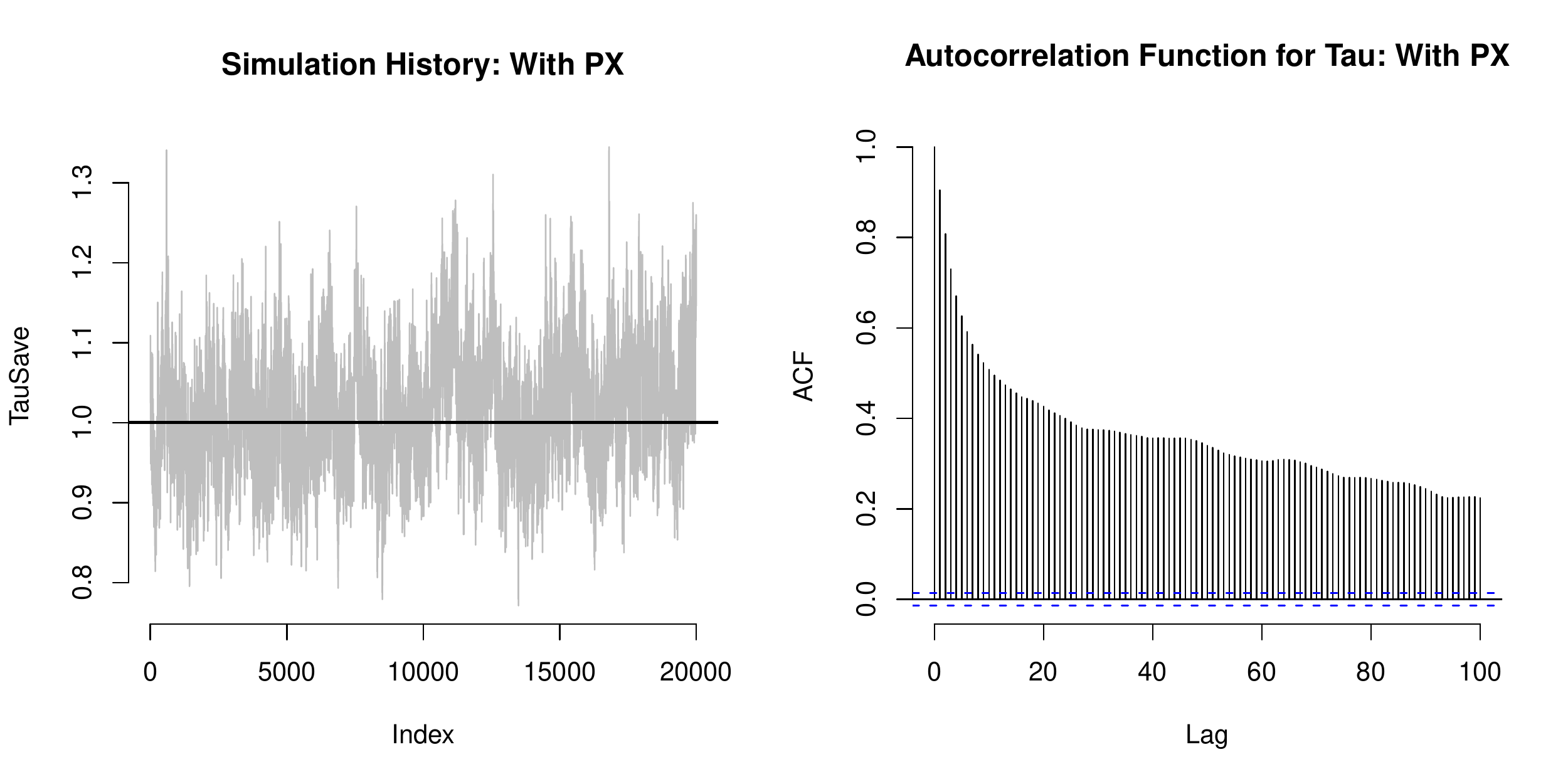}
\caption{\label{fig:locaPXresults} Case 1: $p=1000$, $n=5$, $\sigma=\tau=1$.  Above: Simulation history and autocorrelation plot for $\tau$ in the non-parameter-expanded Gibbs sampler under the horseshoe prior.  Below: the same plots for the parameter-expanded sampler.}
\end{figure}

\begin{figure}
\includegraphics[width=6.0in]{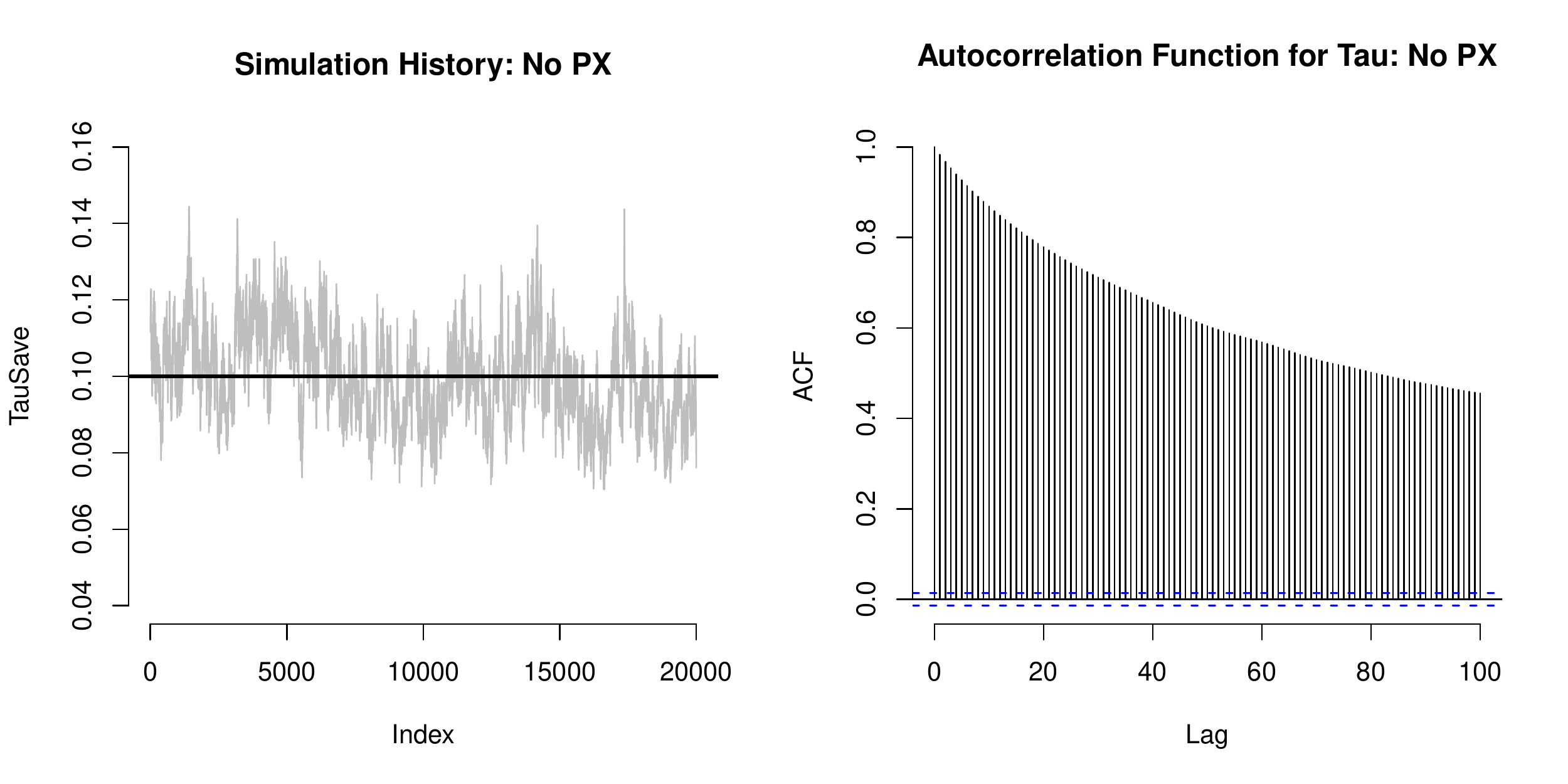}\\
\includegraphics[width=6.0in]{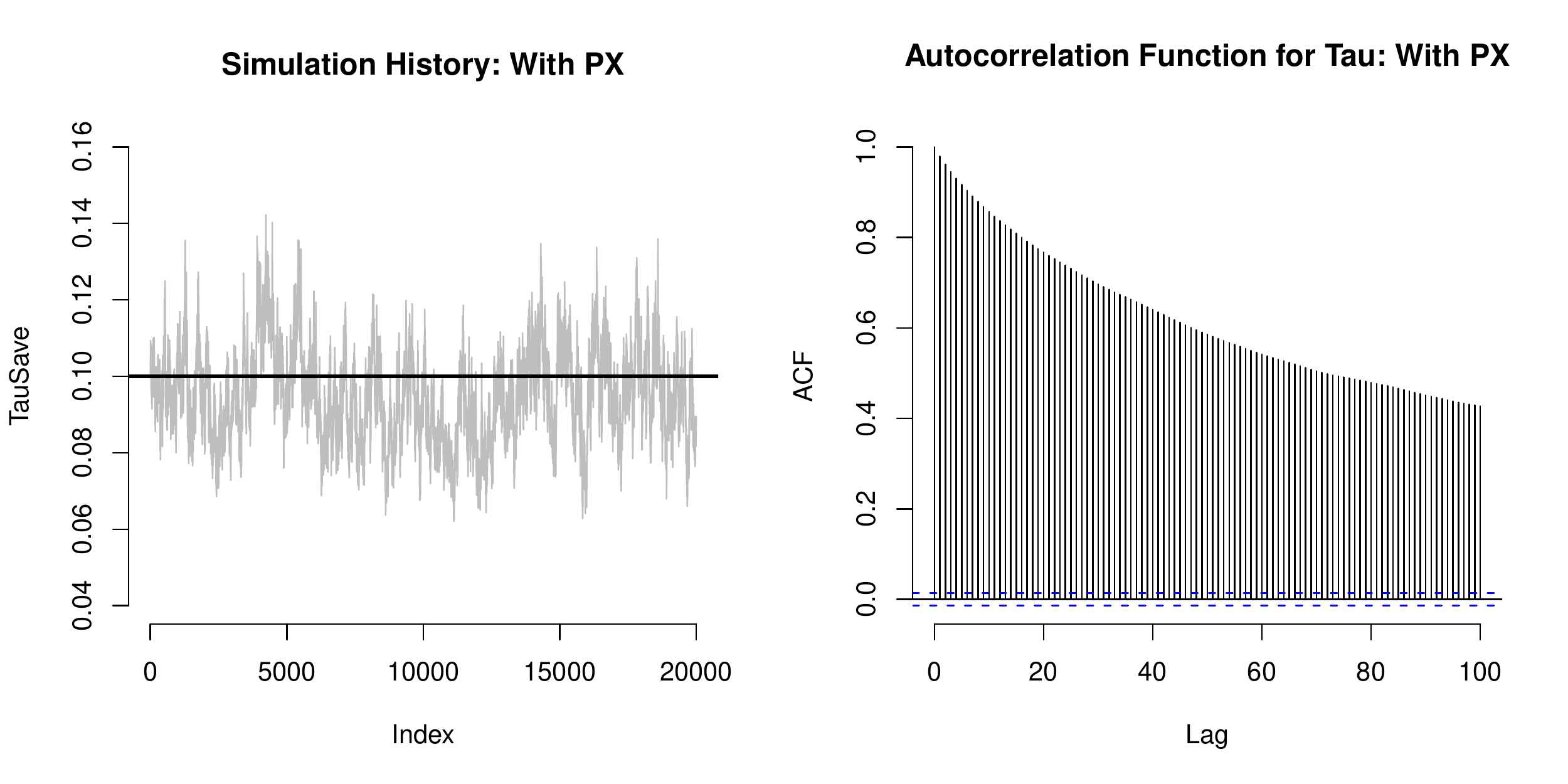}
\caption{\label{fig:locaPXresults-case2} Case 2: $p=2000$, $n=3$, $\sigma=1$, $\tau=0.1$.  Above: Simulation history and autocorrelation plot for $\tau$ in the non-parameter-expanded Gibbs sampler under the horseshoe prior.  Below: the same plots for the parameter-expanded sampler.}
\end{figure}

\begin{figure}
\includegraphics[width=6.0in]{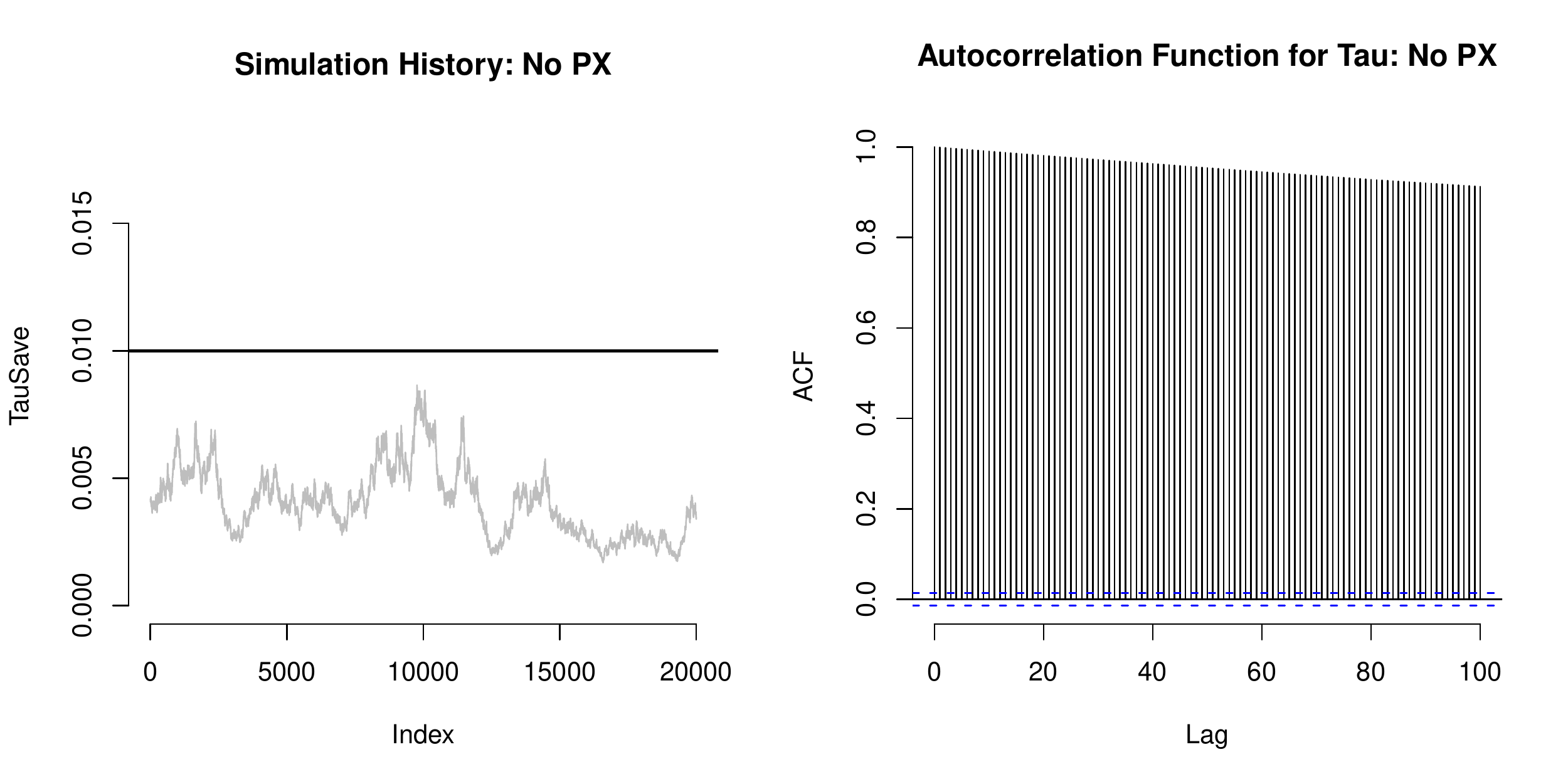}\\
\includegraphics[width=6.0in]{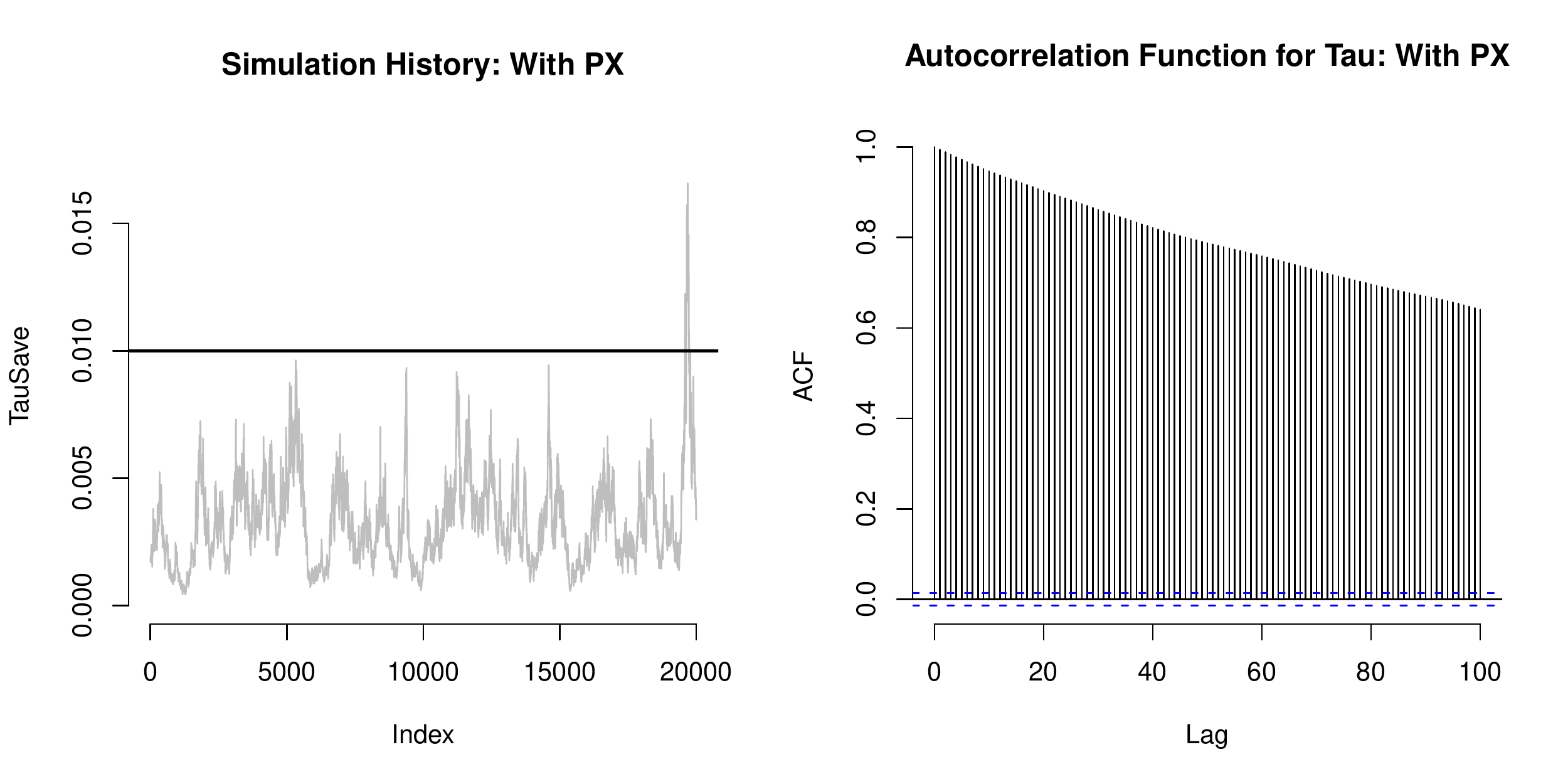}
\caption{\label{fig:locaPXresults-case3} Case 2: $p=5000$, $n=2$, $\sigma=1$, $\tau=0.01$.  Above: Simulation history and autocorrelation plot for $\tau$ in the non-parameter-expanded Gibbs sampler under the horseshoe prior.  Below: the same plots for the parameter-expanded sampler.}
\end{figure}

Figures \ref{fig:locaPXresults}--\ref{fig:locaPXresults-case3} summarize our results for three chosen cases.  In all cases, we ran the PX and non-PX samplers on the same data, simulated from the true model.  We burned-in the samplers for $2\times10^4$ iterations, and saved an additional $2\times10^4$ iterations without any thinning.  The samplers were initialized to the same values, and were provided a stream of pseudo-random numbers from R's default generator starting from the same seed.  Code is provided in the Appendix that allows the reader to replicate these results, and to change the data set or RNG seed.

Unfortunately, it appears that the parameter-expanded sampler offers, at best, only a modest improvement over the non-PX sampler.  In some of the cases explored, the advantage was small but noticeable.  In other cases, there seemed to be virtually no difference.  In no situation that we investigated did we see an improvement anything like that shown for the global-shrinkage-only model (Figure \ref{fig:globalPX}).

This is disappointing, given the importance of these models in Bayesian statistics today.  It is also strikingly different from the case where $p(\beta_j \mid \tau^2)$ is a normal distribution and $\lambda_j \equiv 1$.  We focus on results under the horseshoe prior, but the behavior we witnessed here appears to be quite general for other models, too (e.g.~the Bayesian LASSO).

To quantify the relative efficiency of the two samplers over a variety of different signal-to-noise ratios, we ran the following experiment for all combinations of $\tau \in \{0.01, 0.05, 0.1, 0.5, 1\}$ and $n \in \{1,2,3,4,5\}$.  In all cases, we set $p=1000$ and $\sigma=1$.
\begin{enumerate}
\item Simulate data from the true model for the given configuration of $\tau$ and $n$.
\item Run each sampler (PX and non-PX) for $T=10^5$ iterations after an initial burn-in period of $2 \times 10^4$ iterations.
\item Estimate the effective Monte Carlo sample size as $T_{e} = T/\kappa$ for
$$
\kappa = 1 + 2 \sum_{t=1}^{\infty} \mbox{corr} \left\{ \tau^{(0)}, \tau^{(t)} \right\} \, .
$$
This can be estimated from the MCMC output.
\item Compute the relative efficiency of the PX (P) and non-PX (N) samplers as
$$
r_e = T^{(P)}_{e} / T^{(N)}_{e} \, .
$$
\end{enumerate}

\begin{table}
\begin{center}
\caption{\label{tab:relativeefficiency} Relative efficiency ratios of the PX sampler compared to the non-PX sampler for 20 different configurations of $n$ and $\tau$.  Numbers larger than 1 indicate that the PX sampler is more efficient.}
\vspace{1pc}
\begin{tabular}{rr  r r r r r}
	&& \multicolumn{5}{c}{$\tau$} \\
$n$ && 0.01 & 0.05 & 0.1 & 0.5 & 1 \\
\hline
2 && 13.66 & 1.59 & 1.31 & 1.31 & 1.49 \\
3 && 9.78 & 1.46 & 1.16 & 2.44 & 5.06\\
5 && 9.55 & 1.51 & 0.91 & 5.04 & 11.89\\
10 && 7.36 & 1.02 & 0.79 & 1.31 & 8.70
\end{tabular}
\end{center}
\end{table}

\begin{table}
\begin{center}
\caption{\label{tab:effectivesize} Effective sample size of $10^5$ samples using the parameter-expanded MCMC for 20 different configurations of $n$ and $\tau$.}
\begin{tabular}{rr  r r r r r}
	&& \multicolumn{5}{c}{$\tau$} \\
$n$ && 0.01 & 0.05 & 0.1 & 0.5 & 1 \\
\hline
2 && 1143 & 422 & 554 & 759 & 476 \\
3 && 940 & 537 & 632 & 570 & 313 \\
5 && 936 & 637 & 545 & 570 & 510 \\
10 && 767 & 487 & 463 & 264 & 151 \\
\end{tabular}
\end{center}
\end{table}

For each combination of $\tau$ and $n$, we estimated the relative efficiency for 10 different simulated data sets and averaged the results.  These results are summarized in Table \ref{tab:relativeefficiency}, while Table \ref{tab:effectivesize} shows the average effective sample size for the PX sampler.  From these tables, it is clear that, while the PX sampler usually outperforms the non-PX sampler, it is still highly inefficient.  Here, an MCMC run of length 100,000 tends to yield the equivalent of roughly 100 to 1000 independent draws from the posterior distribution.  These tables, moreover, reflect the performance of the algorithm on a data set with 1000 parameters and no problems introduced by collinear predictors, as might be the case in a regression problem.  For larger problems with collinear predictors, the inefficiencies would likely be much greater.

\section{A final example}

\begin{figure}
\begin{center}
\includegraphics[width=5.0in]{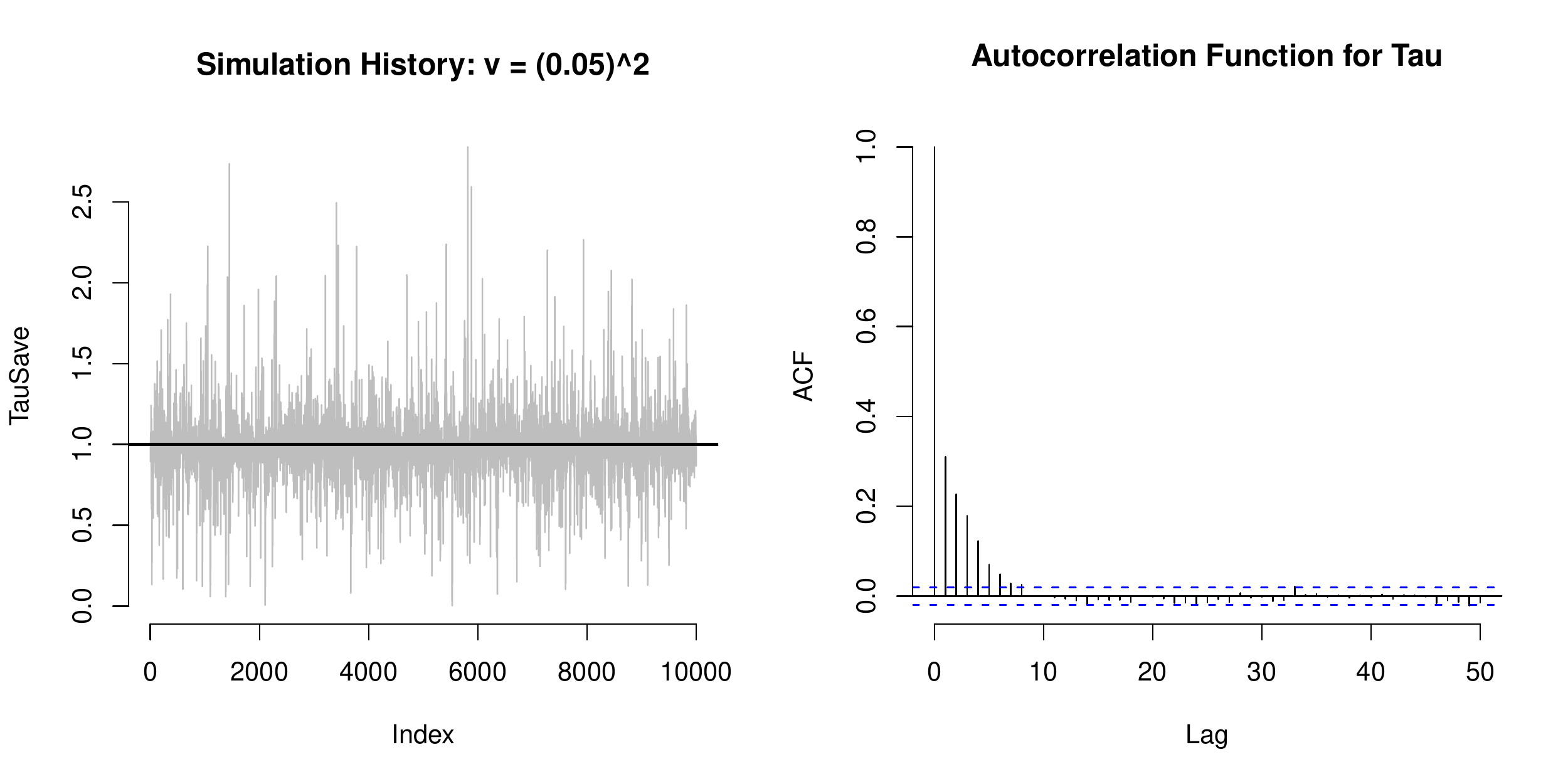}\\
\includegraphics[width=5.0in]{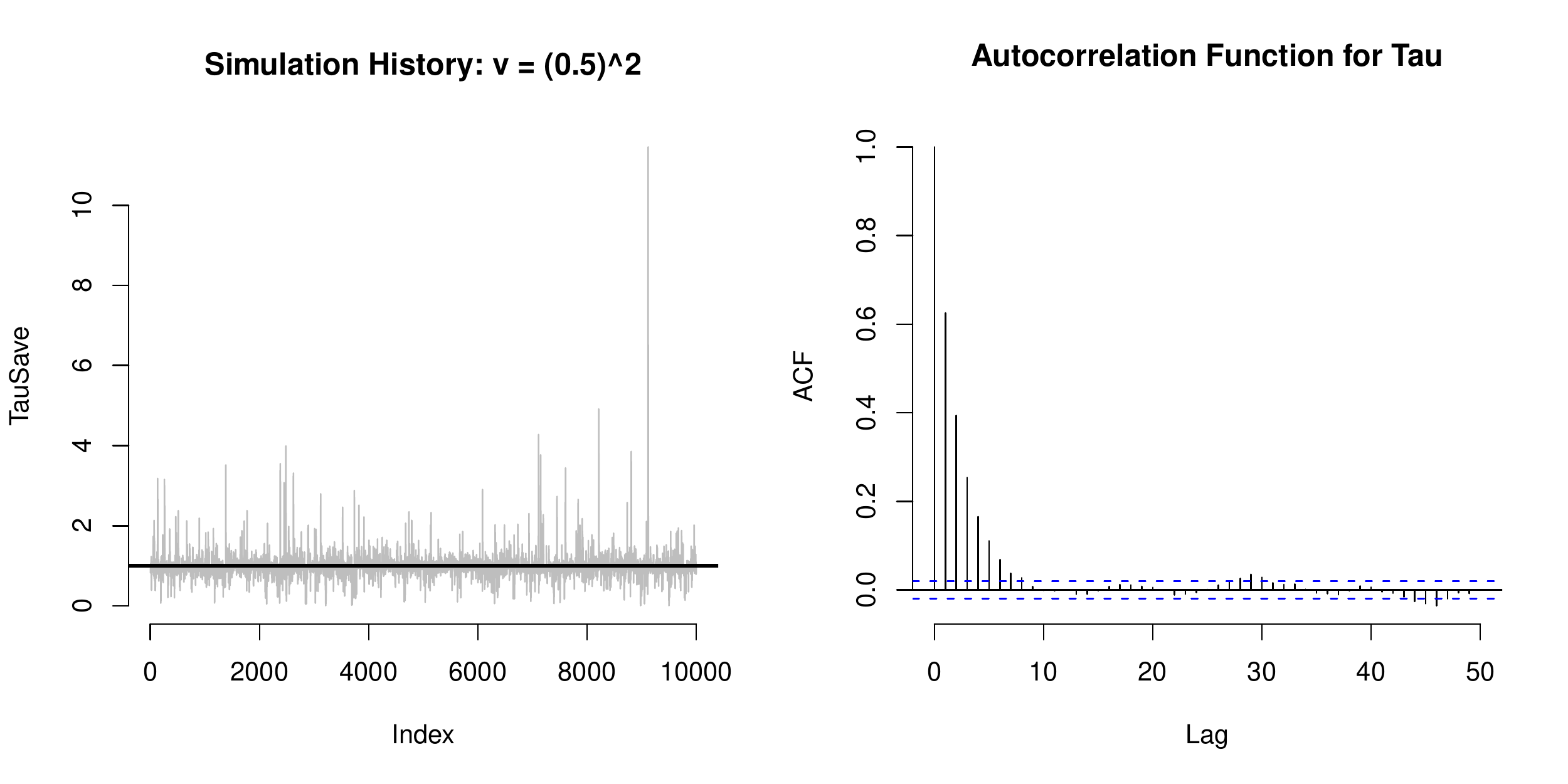}\\
\includegraphics[width=5.0in]{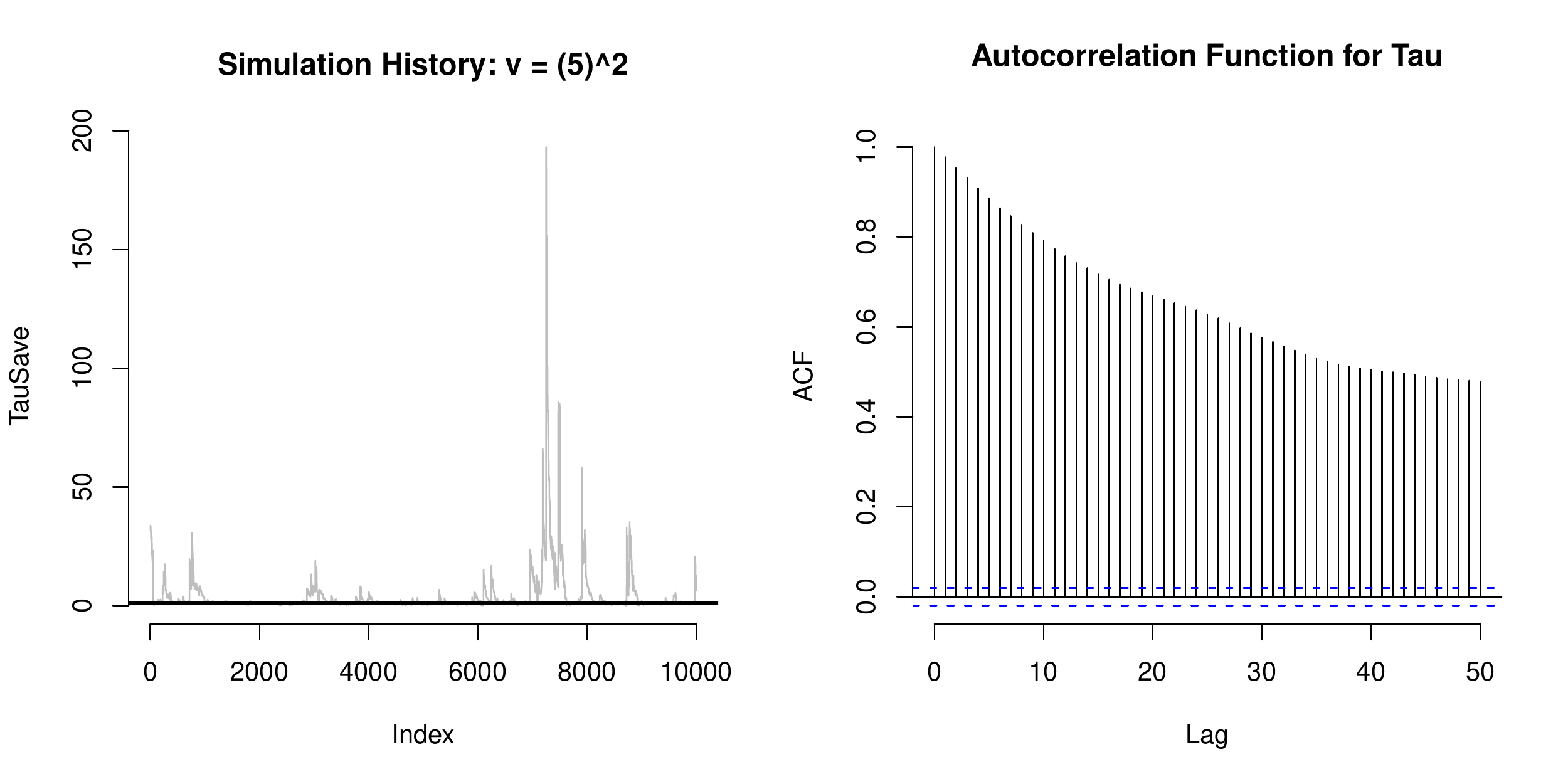}
\caption{\label{fig:localnormal} Three examples of the PX sampler fit to a model where $\lambda_j \sim \N^{+}(1,v)$.  Top: $v=0.05^2$.  Middle: $v = 0.5^2$.  Bottom: $v=5^2$.}
\end{center}
\end{figure}

Intuitively, the failure of PX to offer large gains is due to the presence of the $\lambda_j$'s in the conditionally Gaussian representation for $\beta_j$.  As the directed graphs in Figure \ref{fig:localPXgraph} show, the $\lambda_j$'s fail to be independent of both $\Delta$ and $g$ in the conditional posterior distribution (given $\btheta$ and the data), no matter where they enter the hierarchy.  This induces further autocorrelation in $\tau$, even if the model is structured so that $\Delta$ and $g$ are still conditionally independent of each other.

One final example provides some intuition that, indeed, it does seem to be the $\lambda_j$'s that foul things up.  Suppose that $\lambda_j \sim \N^+(1, v)$ for some pre-specified value of $v$.  When $v$ is small, the $\lambda_j$'s are restricted to lie very near their prior mean of 1.  The resulting model behaves almost exactly like the pure global shrinkage model of (\ref{noPXglobal1})--(\ref{noPXglobal3}).  On the other hand, when $v$ is large, the $\lambda_j$'s are free to vary, and the model may shrink locally rather than just globally.

As we vary $v$ from a large value to a small value, we may study the behavior of the parameter-expanded MCMC.  Figure \ref{fig:localnormal} shows these results for a small experiment $\tau = \sigma = 1$, $n=2$, and $p=1000$.  Notice the increasing degree of autocorrelation as $v$ gets bigger.  This fact suggests that giving the $\lambda_j$'s the freedom to move---as we must do in order to model sparse signals---will inevitably diminish the performance of the parameter-expanded sampler.

\section{Summary}

On balance, parameter expansion appears to be the safest way to handle the global scale parameter $\tau$ in sparse Bayesian models like (\ref{noPXlocal1})--(\ref{noPXlocal4}).  We recommend it as the default option.  But in light of these results, it is difficult to be very excited about this or any other existing technique for fitting such models.  In highly sparse situations, where the posterior for $\tau$ is highly concentrated near zero, MCMC-PX runs of $10^5$ may yield effective sample sizes of only a few hundred draws.  This is precisely the problem that parameter expansion usually solves in hierarchical models.  Its performance on this count simply isn't very encouraging, even if it is usually a bit better than that of the second-best option.

What kind of algorithm could solve the problem?  It is the $\lambda_j$'s, of course, that are at the heart of the matter; they endow the class of conditionally Gaussian models with the ability to handle sparsity, but they make convergence much slower.

Therefore, an algorithm that will solve the problem of autocorrelation in $\tau$ must marginalize over the $\lambda_j$'s in one of two updating steps: either $p(\beta_j \mid \lambda_j, \tau_j, \sigma, \bar{y}_j)$, or $p(\tau \mid \bbeta, \sigma, \Lambda, Y)$.  The update for $\beta_j$, however, depends intimately upon conditional normality, making it a poor candidate for marginalization.

Marginalizing over the $\lambda_j$'s in the update for $\tau$ is tricky.  In most cases it will not even be practical to evaluate the marginal likelihood $p(Y \mid \tau, \sigma)$ without invoking the $\lambda_j$'s.  But there are some families of priors for which it is possible.  In particular, \citet{Polson:Scott:2010c} study the class of hypergeometric inverted-beta priors for a global variance component.  This four parameter generalization of the inverted-beta family yields a form for $p(Y \mid \tau, \sigma)$ that can be expressed in terms of doubly nested hypergeometric series.  These series, however, are sometimes painfully slow to converge, and we have not yet managed to exploit them to produce a ``$\lambda$-marginalized'' Gibbs sampler that is competitive with standard methods.  This remains an active area of research.  As the results of this paper show, a better approach is critically needed.

\singlespacing

\appendix

\section{R code}
\label{app:globalcode}

\subsection{The non-PX sampler}

\begin{footnotesize}
\begin{verbatim}
set.seed(42)
p = 2000
n = 3
TauTrue = 0.1
SigmaTrue = 1

LambdaTrue = abs(rt(p,1))
#LambdaTrue = rep(1,p) 	# Use if you are testing a global-shrinkage model
BetaTrue = rnorm(p,0,SigmaTrue*LambdaTrue*TauTrue)

Y = matrix(0, nrow=p, ncol=n)
for(i in 1:n)
{
	Y[,i] = BetaTrue + rnorm(p,0,SigmaTrue)
}
Ybar = apply(Y,1,mean)

Beta = rep(0,p)
Sigma2 = 1
Sigma = 1
Tau = TauTrue
Lambda = rep(1,p)
Sigma = sqrt(Sigma2)

nmc = 20000
burn = 20000
BetaSave = matrix(0, nrow=nmc, ncol=p)
LambdaSave = matrix(0, nrow=nmc, ncol=p)
TauSave = rep(0, nmc)
Sigma2Save = rep(0, nmc)
Res2 = Y
for(t in 1:(nmc+burn))
{
	if(t %% 1000 == 0) cat("Iteration ",t, "\n")
	
	# First block-update Beta
	a = (Tau^2)*(Lambda^2)
	b = n*a
	s = sqrt(Sigma2*a/{1+b})
	m = {b/{1+b}}*Ybar
	Beta = rnorm(p, m, s)
	Theta = Beta/(Sigma*Lambda)
	
	# Now update Sigma2
	# Jeffreys prior is assumed
	for(i in 1:n)
	{
		Res2[,i] = {Y[,i]^2}/{1+(Tau^2)*{Lambda^2}}
	}
	RSS = sum(Res2)
	Sigma2 = 1/rgamma(1,n*p/2, rate = RSS/2)
	Sigma = sqrt(Sigma2)

	# Now update Tau^2 using slice sampling
	eta = 1/(Tau^2)
	u = runif(1,0,1/(eta + 1))
	ub = (1-u)/u
	a = (p+1)/2
	b = sum(Theta^2)/2
	ub2 = pgamma(ub,a,rate=b)
	u2 = runif(1,0,ub2)
	eta = qgamma(u2,a,rate=b)
	Tau = 1/sqrt(eta)
	
	# Now update Lambda, comment out for global shrinkage only
	Z = Ybar/(Sigma*Theta)
	V2 = 1/rgamma(p,1,rate=(Lambda^2+1)/2)
	num1 = n*V2*(Theta^2)
	den = 1 + num1
	s = sqrt(V2/den)
	m = {num1/den}*Z
	Lambda = rnorm(p,m,s)

	if(t > burn)
	{
		BetaSave[t-burn,] = Beta
		LambdaSave[t-burn,] = Lambda
		TauSave[t-burn] = Tau
		Sigma2Save[t-burn] = Sigma2
	}
}
BetaHat = apply(BetaSave,2,mean)
LambdaHat = apply(abs(LambdaSave),2,mean)
\end{verbatim}
\end{footnotesize}

\subsection{The PX sampler}
\begin{footnotesize}
\begin{verbatim}
set.seed(42)
p = 2000
n = 3
TauTrue = 0.1
SigmaTrue = 1

LambdaTrue = abs(rt(p,1))
#LambdaTrue = rep(1,p) 	# Use if you are testing a global-shrinkage model
BetaTrue = rnorm(p,0,SigmaTrue*LambdaTrue*TauTrue)

Y = matrix(0, nrow=p, ncol=n)
for(i in 1:n)
{
	Y[,i] = BetaTrue + rnorm(p,0,SigmaTrue)
}
Ybar = apply(Y,1,mean)

Beta = rep(0,p)
Sigma2 = 1
Sigma = 1
G = 1
Delta = TauTrue
Tau = abs(Delta)*G
Lambda = rep(1,p)
Sigma = sqrt(Sigma2)

nmc = 20000
burn = 20000
BetaSave = matrix(0, nrow=nmc, ncol=p)
LambdaSave = matrix(0, nrow=nmc, ncol=p)
TauSave = rep(0, nmc)
Sigma2Save = rep(0, nmc)
Res2 = Y
for(t in 1:(nmc+burn))
{
	if(t %% 1000 == 0) cat("Iteration ",t, "\n")
	
	# First block-update Beta
	a = (Tau^2)*(Lambda^2)
	b = n*a
	s = sqrt(Sigma2*a/{1+b})
	m = {b/{1+b}}*Ybar
	Beta = rnorm(p, m, s)
	Theta = Beta/(Sigma*Delta*Lambda)
	
	# Now update Sigma2
	# Jeffreys prior is assumed
	for(i in 1:n)
	{
		Res2[,i] = {Y[,i]^2}/{1+(Tau^2)*{Lambda^2}}
	}
	RSS = sum(Res2)
	Sigma2 = 1/rgamma(1,n*p/2, rate = RSS/2)
	Sigma = sqrt(Sigma2)

	# Now update Tau^2
	# Method 2: parameter expansion
	{
		G = 1/sqrt(rgamma(1,(p+1)/2, rate = (1+sum(Theta^2))/2))
		Z = Ybar/(Sigma*Theta*Lambda)
		a = n*(Lambda*Theta)^2
		b = sum(a)
		s2 = 1/(1+b)
		m = {s2}*sum(a*Z)
		Delta = rnorm(1,m,sqrt(s2))
		Tau = abs(Delta)*G
	}
	
	# Now update Lambda, comment out for global shrinkage only
	Z = Ybar/(Sigma*Delta*Theta)
	V2 = 1/rgamma(p,1,rate=(Lambda^2+1)/2)
	num1 = n*V2*((Delta*Theta)^2)
	den = 1 + num1
	s = sqrt(V2/den)
	m = {num1/den}*Z
	Lambda = rnorm(p,m,s)

	if(t > burn)
	{
		BetaSave[t-burn,] = Beta
		LambdaSave[t-burn,] = Lambda
		TauSave[t-burn] = Tau
		Sigma2Save[t-burn] = Sigma2
	}
}
BetaHat = apply(BetaSave,2,mean)
LambdaHat = apply(abs(LambdaSave),2,mean)
\end{verbatim}
\end{footnotesize}

\singlespace
\bibliographystyle{abbrvnat}
\bibliography{masterbib}

\begin{thebibliography}{18}
\providecommand{\natexlab}[1]{#1}
\providecommand{\url}[1]{\texttt{#1}}
\expandafter\ifx\csname urlstyle\endcsname\relax
  \providecommand{\doi}[1]{doi: #1}\else
  \providecommand{\doi}{doi: \begingroup \urlstyle{rm}\Url}\fi

\bibitem[Bae and Mallick(2004)]{bae:mallick:2004}
K.~Bae and B.~Mallick.
\newblock Gene selection using a two-level hierarchical {B}ayesian model.
\newblock \emph{Bioinformatics}, 20\penalty0 (18):\penalty0 3423--30, 2004.

\bibitem[Berger(1980)]{BergerAnnals1980}
J.~O. Berger.
\newblock A robust generalized {B}ayes estimator and confidence region for a
  multivariate normal mean.
\newblock \emph{The Annals of Statistics}, 8\penalty0 (4):\penalty0 716--761,
  1980.

\bibitem[Carlin and Polson(1991)]{carlin:polson:1991}
B.~P. Carlin and N.~G. Polson.
\newblock Inference for nonconjugate {B}ayesian models using the gibbs sampler.
\newblock \emph{The Canadian Journal of Statistics}, 19\penalty0 (4):\penalty0
  399--405, 1991.

\bibitem[Carvalho et~al.(2010)Carvalho, Polson, and
  Scott]{Carvalho:Polson:Scott:2008a}
C.~M. Carvalho, N.~G. Polson, and J.~G. Scott.
\newblock The horseshoe estimator for sparse signals.
\newblock \emph{Biometrika}, 97\penalty0 (2):\penalty0 465--80, 2010.

\bibitem[Damien et~al.(1999)Damien, Wakefield, and Walker]{damien:etal:1999}
P.~Damien, J.~C. Wakefield, and S.~G. Walker.
\newblock Bayesian nonconjugate and hierarchical models by using auxiliary
  variables.
\newblock \emph{J.~R.~Stat.~Soc.~Ser.~B, Stat.~Methodol.}, 61:\penalty0
  331--44, 1999.

\bibitem[Figueiredo(2003)]{figueiredo:2003}
M.~Figueiredo.
\newblock Adaptive sparseness for supervised learning.
\newblock \emph{IEEE Transactions on Pattern Analysis and Machine
  Intelligence}, 25\penalty0 (9):\penalty0 1150--9, 2003.

\bibitem[Gelman(2006)]{gelman:2006}
A.~Gelman.
\newblock Prior distributions for variance parameters in hierarchical models.
\newblock \emph{Bayesian Anal.}, 1\penalty0 (3):\penalty0 515--33, 2006.

\bibitem[Gramacy and Pantaleo(2010)]{gramacy:pantaleo:2009}
R.~Gramacy and E.~Pantaleo.
\newblock Shrinkage regression for multivariate inference with missing data,
  and an application to portfolio balancing.
\newblock \emph{Bayesian Analysis}, 5\penalty0 (2), 2010.

\bibitem[Griffin and Brown(2005)]{griffin:brown:2005}
J.~Griffin and P.~Brown.
\newblock Alternative prior distributions for variable selection with very many
  more variables than observations.
\newblock Technical report, University of Warwick, 2005.

\bibitem[Hans(2009)]{hans:2008}
C.~M. Hans.
\newblock {B}ayesian lasso regression.
\newblock \emph{Biometrika}, 96\penalty0 (4):\penalty0 835--45, 2009.

\bibitem[Hans(2010)]{hans:2010}
C.~M. Hans.
\newblock Model uncertainty and variable selection in {B}ayesian lasso
  regression.
\newblock \emph{Statistics and Computing}, 20:\penalty0 221--9, 2010.

\bibitem[Jeffreys(1961)]{jeffreys1961}
H.~Jeffreys.
\newblock \emph{Theory of Probability}.
\newblock Oxford University Press, 3rd edition, 1961.

\bibitem[Park and Casella(2008)]{park:casella:2008}
T.~Park and G.~Casella.
\newblock The {B}ayesian lasso.
\newblock \emph{Journal of the American Statistical Association}, 103\penalty0
  (482):\penalty0 681--6, 2008.

\bibitem[Polson and Scott(2010)]{Polson:Scott:2010c}
N.~G. Polson and J.~G. Scott.
\newblock On the half-{C}auchy prior for a global scale parameter.
\newblock Technical report, University of Texas at Austin, 2010.

\bibitem[Strawderman(1971)]{strawderman:1971}
W.~Strawderman.
\newblock Proper {B}ayes minimax estimators of the multivariate normal mean.
\newblock \emph{The Annals of Statistics}, 42:\penalty0 385--8, 1971.

\bibitem[Tibshirani(1996)]{tibshirani1996}
R.~Tibshirani.
\newblock Regression shrinkage and selection via the lasso.
\newblock \emph{J. Royal. Statist. Soc B.}, 58\penalty0 (1):\penalty0 267--88,
  1996.

\bibitem[Tipping(2001)]{tipping:2001}
M.~Tipping.
\newblock Sparse {B}ayesian learning and the relevance vector machine.
\newblock \emph{Journal of Machine Learning Research}, 1:\penalty0 211--44,
  2001.

\bibitem[van {D}yk and Meng(2001)]{vandyk:meng:2001}
D.~van {D}yk and X.~L. Meng.
\newblock The art of data augmentation (with discussion).
\newblock \emph{Journal of Computational and Graphical Statistics},
  10:\penalty0 1--111, 2001.

\end{thebibliography}

\end{document}